\documentclass[%
reprint,
superscriptaddress,
amsmath,amssymb,
prl,
]{revtex4-2}
 \usepackage[utf8]{inputenc}
\usepackage{graphicx}
\usepackage{float}
\usepackage{xcolor}
\usepackage{physics}
\usepackage{slashed}
\usepackage[utf8]{inputenc}
\usepackage{cancel}
\graphicspath{ {./figures/} }

\begin{document}

\title{Localization of Active Particles on Random Arrays of Parallel Filaments}
\author{Owen Santoso}
\affiliation{Department of Physics,
	University of California San Diego,
	La Jolla, CA 92093}
\author{Elena F. Koslover}
\email{ekoslover@ucsd.edu}
\affiliation{Department of Physics,
	University of California San Diego,
	La Jolla, CA 92093}

\begin{abstract}
    Quenched disorder in the environment can fundamentally alter transport dynamics in both active and passive systems. We explore how disordered arrays of filaments govern the distribution of intermittently moving particles which switch between diffusive and processive transport. Motivated by the mixed-polarity arrangements of parallel microtubules observed in mammalian dendrites, we show that such arrays tend to result in localization of particles at regions of convergent filament orientation.
    In the rapid attachment-detachment limit, the disordered system can be described by a noisy one-dimensional effective energy landscape, whose structure is approximated by a random walk. The depth and width of wells on this landscape are expressed as a function of the transport kinetics and system geometry. Localization is shown to be strongest at intermediate run-lengths, where biased transport persists long enough to sense the quenched filament polarity but not so long as to facilitate escape from local traps.
    These results demonstrate robust localization of particles moving on random filament networks, highlighting the emergent spatial organization that arises from an interplay of active transport and quenched disorder.
\end{abstract}

\maketitle

Random motion through a disordered environment~\cite{hughes1996random} appears in a variety of physical systems, including glasses~\cite{debenedetti2001supercooled},
porous media~\cite{bouchaud1990anomalous}, crowded cytoplasm~\cite{destrian2026cytoplasmic}, and proteins sliding along DNA~\cite{lu2021search}.
Such quenched disorder can greatly reduce propagation and relaxation timescales, leading to subdiffusion~\cite{bouchaud1990anomalous,chepizhko2013diffusion} and ultra-slow transport in one-dimensional systems~\cite{sinai1983limiting}.
Networks with quenched disorder tend to exhibit eigenvector localization, wherein modes with widely different decay times  are constrained to small regions of space~\cite{goltsev2012localization,chaudhuri2014diversity}. This results in patchy relaxation dynamics, with individual regions reaching metastable states long before communication between regions can occur. The steady-state distribution may also localize, converging to the lowest energy well in a passive system  or to topological sinks in an active system with convergent flows~\cite{bouchaud1990anomalous}. Eigenmode localization plays an important role in understanding a variety of complex network behaviors, such as epileptic seizures in neural networks~\cite{zhang2017suppression}, population clustering in a fluctuating  ecology~\cite{nelson1998non}, and epidemic spread through heterogeneous populations~\cite{wei2020susceptible}.

Inside a living cell, the localization of organelles, proteins, and RNA allows the establishment of functionally distinct subcellular regions~\cite{van2016right}. Such localization is crucial in specialized neuronal cells, where functional components must be positioned in distal regions far from the cell body~\cite{misgeld2017mitostasis}.
Intracellular localization can be achieved through a combination of motor-driven directed transport along organized cytoskeletal highways~\cite{burute2019cellular,kapitein2011way}, binding to fixed cellular structures~\cite{koslover2024searching,maza2019intrinsic}, and cytoplasmic fluid flow patterns~\cite{illukkumbura2020patterning}.

Organelles such as endosomes~\cite{yap2022spatial,chen2015memoryless} and lysosomes~\cite{goo2017activity} are known to undergo multimodal motion, alternating between local diffusive exploration, and directed runs that switch between cytoskeletal tracks~\cite{balint2013correlative}. The organization and polarization of the cytoskeleton plays an important role in subcellular positioning of these particles~\cite{burute2019cellular,kapitein2011way}.
Microtubule density has been shown to modulate vesicle pausing and directional switching rates {\em in vivo}~\cite{yogev2016microtubule}, while {\em in vitro} actin arrays demonstrate particle accumulation at regions of vanishing net polarity~\cite{richard2019active}.
In disordered filament arrays, local regions where filament orientations converge can trap motor-driven particles.
Such traps have been observed in reconstituted  actin networks, where motor-driven cargos  enter long-lived localized cycling states that lead to emergent glassy dynamics~\cite{scholz2016cycling}. Mathematical models show that individual convergent regions have exponentially slow escape times~\cite{sarpangala2024tunable} and that trapping in random filament arrays can slow down the rate at which intermittently moving particles explore space, find targets, or encounter one another~\cite{ando2015cytoskeletal,mlynarczyk2019first,teryoshin2026encounter}.

Unpolarized arrays of parallel microtubules are found in the proximal region of mammalian dendrites~\cite{kapitein2010mixed,katrukha2021quantitative}, where they have been suggested to serve the functional purpose of broadly distributing vesicular cargos~\cite{kapitein2011way}. Although cytoskeletal filaments can be highly dynamic, neuronal projections contain microtubule bundles that remain stable over long timescales~\cite{baas2016stability,katrukha2021quantitative}. Thus, dendritic microtubule arrays serve as an exemplar system for  examining how quenched filament disorder governs the spatial organization of intermittently active particles.

Here we show that transport on parallel filament bundles can be approximated by an effective potential landscape whose ruggedness is determined by a combination of transport kinetics, filament length, and density. We demonstrate that localization rather than wide dispersion of cargos is expected under a broad range of parameters, and is most pronounced for intermediate cargo run lengths. These results go beyond classic theories of transport on random landscapes~\cite{sinai1983limiting,golosov1984localization,bouchaud1990anomalous,hughes1996random} to elucidate how cytoskeletal structure shapes the distribution of motor-driven cargos in living cells.

\begin{figure}[t!]  % or [!t], [!b], [p], etc.
    \centering
    \includegraphics[width=\columnwidth]{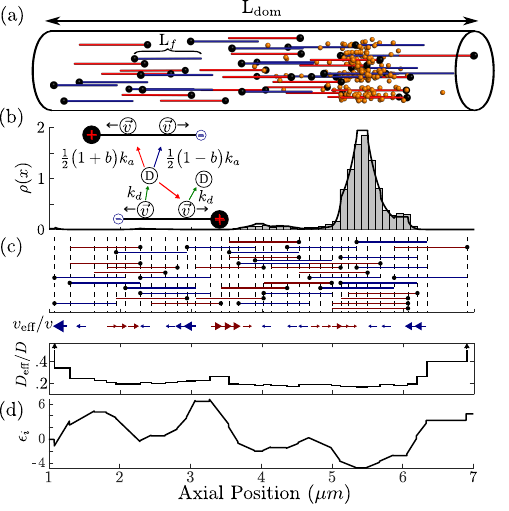}
    \caption{
        (a) Simulation snapshot of biased particles (orange) navigating a parallel filament array, with filament plus ends pointing right (red) or left (blue). Only the axial dimension is resolved in the simulations.
        (b) Inset: schematic of transitions between motility states. Histogram: steady-state distribution from simulations, Line: analytic distribution from Eq.~\ref{eq:landscape}. (c) Region decomposition of the filament array, showing effective velocity (arrows) and diffusivity (black curve) in each region. (d) Energy landscape corresponding to the filament configuration. Parameters used: $L_f = 1\mu \text{m}, n = 7, L_\text{dom} = 7\mu\text{m}, v = 1\mu \text{m}, D = 0.1\mu \text{m}^2,k_a = k_d = 100s^{-1}, b=1$.
    }
    \label{fig:Simulation}
\end{figure}

We consider the motion of particles that switch between passive diffusion and active motion along a fixed array of parallel filaments in a long tubular domain of length $L_\text{dom}$ (Fig.~\ref{fig:Simulation}a). The domain is assumed to be so narrow that diffusive particles can simultaneously access all filaments passing through a given axial position, so that particle motion can be approximated as one-dimensional along the axial coordinate ($x$). Filaments possess intrinsic polarity, with distinct plus and minus ends. A total of $N_f$ filaments, each of of length $L_f$, are uniformly scattered throughout the domain, with average number per cross-section $n= N_f L_f/L_\text{dom}$.
Each filament is assigned a plus and minus end at random, producing a mixed-polarity array.

Detached particles undergo passive diffusion with diffusivity $D$ and a constant rate of attachment $k_a$ per filament in the cross-section (Fig.~\ref{fig:Simulation}b, inset). Upon attaching to a  filament, the particle selects a direction of motion with the bias parameter $b\in[-1,1]$ defining the relative probability of walking towards the plus versus the minus end.
Bias $b=1$ represents particles that always walk to the plus end, while $b=0$ represents unbiased particles equally likely to move in either direction.
Attached particles move along an individual filament with speed $v$ and detach with constant rate $k_d$. Particles that reach the end of a filament remain stationary until detachment occurs. To switch the direction of processive motion, particles must first detach and then reattach to another filament.
Stochastic simulations of this intermittent transport model indicate that  steady-state distributions tend to be highly localized, with particles accumulating in traps formed by convergent filaments (Fig.~\ref{fig:Simulation}b).

To analyze how particle distribution depends on kinetic and geometric parameters, we first break up the domain into a discrete sequence of regions separated by boundaries wherever filaments start or end (Fig.~\ref{fig:Simulation}c). Each region has length $\ell_i$ and contains $n_i = n_i^+ + n_i^-$ filaments, where $n_i^\pm$ is the number of filaments pointing in each direction. Within the region, particles switch between motility states at fixed rates, with total rate $n_ik_a$ of entering the active state. The local directional bias for active motion is characterized by the asymmetry parameter $\alpha_i = b(n^+_{i} - n^-_{i})/n_{i}$.

To simplify the mathematical analysis, we make a crucial  `single-layer' approximation~\cite{sarpangala2024tunable}, where the attachment and detachment rates are assumed to be very fast compared to the timescale of traversing a filament ($k_a, k_d \gg v/L_\text{MT}$). In this rapid-exchange limit, the particle experiences an effective advection velocity $v_\text{eff,i}$ and effective diffusivity $D_\text{eff,i}$ within each region, which can be derived from the long-time drift and dispersion of intermittently switching particles~\cite{supplemental} to yield:
\begin{subequations}\label{vDeff}
    \begin{align}
        v_{\mathrm{eff},i} & = v \alpha_i f_{a,i}
        \label{eq:veff}                                                                                                                          \\
        D_{\mathrm{eff},i} & = D f_{d,i}  + \frac{v^2 f_{a,i}}{k_d} \left[1-\alpha_i^2 + \left(f_{d,i} \alpha_i\right)^2\right]. \label{eq:Deff}
    \end{align}
\end{subequations}
Here $f_{d,i} = k_d/(n_ik_{a} + k_d), f_{a,i} = 1 - f_{d,i}$ are the steady-state fractions of time spent in the passive and active state, respectively.

The overall density $\rho_i(x)$ of particles at axial position $x$ within each region is found from the steady-state solution of an advection-diffusion equation, with a discontinuity at each boundary:
\begin{subequations}
    \begin{align}
        \frac{d\rho_i}{dt} = -v_\text{eff,i} \rho_i'(x) + D_\text{eff,i} \rho_i''(x) = 0 \label{eq:rho} \\
        (k_d + n_{i}k_a) \rho_{i-1}(\ell_{i-1}) = (k_d + n_{i-1}k_a) \rho_i(0). \label{eq:rhobound}
    \end{align}
\end{subequations}
The discontinuity is set such that the density of diffusive particles ($f_{d,i} \rho_i$) is continuous across the boundaries. For this one-dimensional system, the steady-state solution~\cite{supplemental} can be expressed in terms of an effective energy function $\epsilon_i(x)$, with
\begin{subequations}
    \begin{align}
        \rho_i(x)     & = \frac{1}{\mathcal{N}}e^{-\epsilon_i(x)} \label{eq:rho}                                                                        \\
        \epsilon_i(x) & = -\sum_{j=1}^{i-1} \text{Pe}_j - \frac{v_\text{eff,i}}{D_\text{eff,i}} x - \log\left[\frac{k_d + n_i k_a}{k_d + n k_a}\right],
        \label{eq:rhoE}
    \end{align}
    \label{eq:landscape}
\end{subequations}
where $\mathcal{N}$ is a normalization constant across the whole domain and $\text{Pe}_j = v_j \ell_j/D_j$ is the P\'eclet number describing the balance of drift and diffusion within each region. The first two terms in Eq.~\ref{eq:rhoE} encompass the biased motion of particles towards convergent regions. The logarithmic term describes an entropic effect where particles are more likely to be found in regions with many available filaments, even if there is no active motion.
The distribution given by Eq.~\ref{eq:rho} matches to that obtained via stochastic simulations (Fig.~\ref{fig:Simulation}b), with wells in the effective energy landscape corresponding to regions of local particle accumulation (Fig.~\ref{fig:Simulation}d).

\begin{figure}[t!] % or [!t], [!b], [p], etc.
    \centering
    \includegraphics[width=\columnwidth]{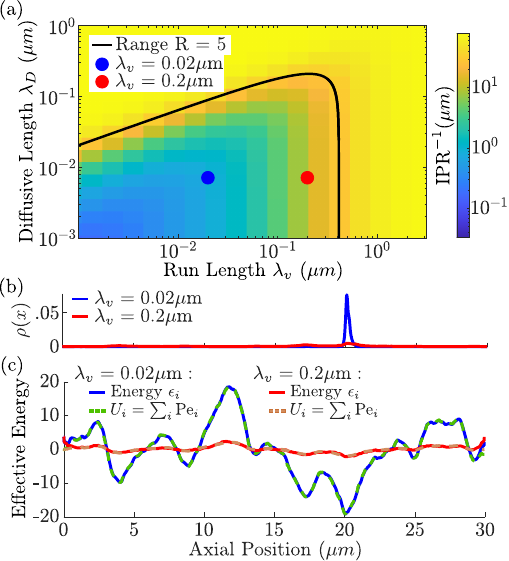}
    \caption{ (a) Steady-state localization width (IPR$^{-1}$, color), plotted as a function of diffusive length $\lambda_D$ and run length $\lambda_v$. Black curve tracks values where the energy landscape range (Eq.~\ref{eq:range}) reaches $R=5$, marking a transition to tight localization.
        (b) Steady-state distribution of particles on a sample filament configuration for two different run lengths ($\lambda_v = 0.02\mu\text{m}, 0.2\mu\text{m}$). (c)
        Energy landscapes ($\epsilon_i$) corresponding to the distributions in (b) are shown in blue and red. The potential function $U_i$ (green and brown) provides a smoothed estimate of the landscape.  Parameters used: $L_f = 1\mu \text{m}, n = 40, k_a = k_d = 100s^{-1}, b=1$ and $L_\text{dom} = 100\mu\text{m}$ in (c). IPR$^{-1}$ color plot is averaged over 200 configurations.
    }
    \label{fig:landscape}
\end{figure}

Localization  on a particular landscape can be described via the inverse participation ratio (IPR)~\cite{goltsev2012localization,nelson2024nonreciprocity}, defined as $\text{IPR} = \sum_i \int_0^{\ell_i} \rho_i^2(x) dx $.
The inverse of this quantity ($\text{IPR}^{-1}$) has units of length and approximately describes the support of the distribution function. A perfectly uniform distribution has $\text{IPR}^{-1} = L_\text{dom}$, while a tightly localized distribution will have $\text{IPR}^{-1} \rightarrow 0$.
As shown in Fig.~\ref{fig:landscape}, localization depends on the kinetic parameters defining particle motion. Tight localization occurs when the diffusive length $\lambda_D \equiv \sqrt{D/(n k_a)}$ is small and the run length $\lambda_v \equiv v/k_d$ has intermediate values. Conceptually, active runs have counterbalancing effects --  making particles more sensitive to filament orientation, but also facilitating their ability to escape from traps.

To unravel the features that govern particle trapping, we note that this system is analogous to the classic problem of motion along a one-dimensional random landscape~\cite{sinai1983limiting,zwanzig1988diffusion}
and proceed to examine how the shape of this landscape is determined by particle kinetics and filament geometry.
We consider the limit where filaments are dense ($n\gg 10$) and particles spend most of their time in the active state ($f_d\ll 1$). This approximation explicitly neglects the edges of the domain where only a few filaments are present, and we focus specifically on localization within the bulk where filament density is high. For large values of $n$, the individual filament counts within each region ($n_i$) should be tightly peaked around their average value. The energy landscape is then accurately approximated by the potential term formed from the accumulation of P\'eclet numbers: $U_i = -\sum_{j=1}^{i-1}\text{Pe}_j$ (Fig.~\ref{fig:landscape}c).

The P\'eclet numbers can be expressed in terms of the average  run length $\lambda_v$ and diffusive length $\lambda_{D}$, giving
\begin{equation}
    \begin{split}
        Pe_i \approx \frac{\lambda_v \alpha_i \ell_i }{\lambda_{D}^2 + \lambda_v^2 \left[1- \alpha_i^2\right]}.
    \end{split}
\end{equation}
The number of positively oriented filaments in each region ($n^+_i$) is a binomially distributed random variable with mean $n/2$ and variance $n^2/4$. The asymmetry parameter $\alpha_i$ thus has mean $0$ and variance $b^2/n^2$. The region lengths $\ell_i$ can be approximated by uniformly scattering $2N_{f}$ filament end-points in the domain. For a large number of filaments,  the separation between nearest-neighbor points should have mean $\ell = L_\text{dom}/(2N_f) = L_f/(2n)$ and variance $\ell^2$~\cite{supplemental}. The P\'eclet number in each region then has mean $0$ and standard deviation:
\begin{equation}
    \begin{split}
        \sigma = \frac{\lambda_v}{\lambda_D^2+\lambda_v^2} \frac{bL_f}{\sqrt{2n^3}}.
        \label{eq:sigPeMain}
    \end{split}
\end{equation}

While we would expect the region lengths to be independent of each other, the asymmetry $\alpha_i$ should be correlated between consecutive regions. We can express $\alpha_i = (b/n_i)\sum_{j=1}^{n_i} w_j$, where each $w_j = \pm 1$ describes the orientation of a single filament. Moving across the domain, each time one filament ends and another begins, one of the original $w_j$ is removed and a new, uncorrelated value is added. The resulting covariance in $\alpha$ decreases linearly with the region count~\cite{supplemental}, giving the  covariance function $\left<\text{Pe}_0 \text{Pe}_i\right> = \frac{1}{2}\sigma^2 ( 1 - \frac{i}{2n})$ for $ 0< i \leq 2n$.
On average, a span of $2n$ consecutive regions corresponds to one filament length $L_f$, setting the correlation length of the individual P\'eclet numbers.
By taking the continuum limit of very small regions and integrating the covariance function~\cite{supplemental}, we can compute the typical root-mean-squared step size $s$ in the potential landscape over the course of one correlation length, obtaining %$s \approx 2n\sigma/\sqrt{3}$.
\begin{equation}
    \begin{split}
        s = \sqrt{\left<\left(\frac{1}{\ell}\int_0^{L_f} \text{Pe}(x)dx\right)^2\right>} \approx \frac{2n\sigma}{\sqrt{3}}.
    \end{split}
    \label{eq:stepsize}
\end{equation}

We can then approximate the potential landscape $U$ as a discrete random walk with $N = L_\text{dom}/L_f$ independent steps of typical size $s$.
The average range $R$ (maximum to minimum difference) of the landscape is estimated using the standard expression for a random walk~\cite{hughes1996random}:
\begin{equation}
    \begin{split}
        R = \sqrt{\frac{8}{\pi}} s \sqrt{N} \approx \sqrt{\frac{16}{3\pi}} \left(\frac{b\lambda_v}{\lambda_D^2 + \lambda_v^2}\right) \sqrt{\frac{L_\text{dom}L_f}{n}}.
    \end{split}
    \label{eq:range}
\end{equation}
As shown in Fig.~\ref{fig:landscape}, tight localization of particle distributions requires this range to be sufficiently large, indicating a deep well in the energy landscape.

\begin{figure}  % or [!t], [!b], [p], etc.
    \centering
    \includegraphics[width=\columnwidth]{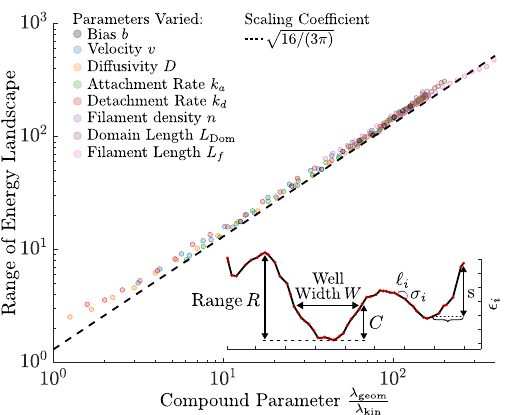}
    \caption{
        Depth of energy landscape collapses to a universal curve as a function of the ratio of length scales $\lambda_{\text{geom}}/\lambda_\text{kin}$. Inset illustrates the landscape range $R$, well width $W$, and correlated step size $s$ for a truncated energy landscape, with red dots marking region boundaries.
        Dots show range of the bulk landscape $\epsilon$, averaged over 200 sampled configurations, while kinetic and geometric parameters are varied. Dashed line gives analytic prediction from Eq.~\ref{eq:range}.
        Default parameter values: $b = 1, v = 1\mu \text{m}, D = 0.1\mu^2\text{m}/s, k_a = k_d = 100s^{-1}, n = 40, L_{\text{dom}} =100 \mu \text{m}, L_f = 1\mu \text{m}$. Edge regions within distance $L_f$ of the domain boundaries are excluded from the calculation.
    }
    \label{fig:range}
\end{figure}

The expression in Eq.~\ref{eq:range} highlights the key conditions for localization. It includes two length scales: a kinetic length $\lambda_\text{kin} = (\lambda_v^2 + \lambda_D^2)/(b\lambda_v)$, and a geometric length $\lambda_\text{geom} = \sqrt{L_\text{dom}L_f/n}$, with localization occuring when $\lambda_\text{geom}/\lambda_\text{kin}$ is sufficiently large.
Figure~\ref{fig:range} demonstrates that the landscape range collapses to a universal curve as a function of this ratio of length-scales. For very small values of the ratio, some deviation  is observed because the entropic term describing the variability in filament density $n_i$ begins to dominate the disorder. In this regime, particles are preferentially found in regions with more filaments, resulting in an overall broad distribution throughout the domain.

The kinetic length $\lambda_\text{kin}$ is minimized at intermediate values of the run length $\lambda_v$. Very small run lengths imply that most of the spatial motion occurs in the diffusive state and the system is insensitive to the biased drift. By contrast, very large run lengths increase the active dispersion, raising $D_\text{eff}$. The geometric length is large when the domain is long (making it more likely for a deeply convergent region to develop), and the number of filaments per cross-section $n$ is not too large (allowing for greater fluctuations in the local directional bias).
The filament length $L_f$ must also be long, to enable sufficiently large correlated steps in the potential landscape.

\begin{figure}  % or [!t], [!b], [p], etc.
    \centering
    \includegraphics[width=\columnwidth]{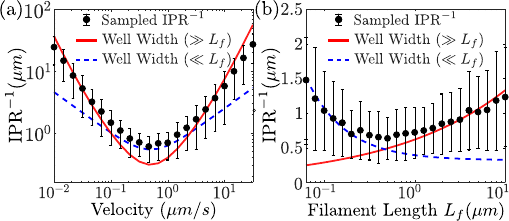}
    \caption{
        Localization width, quantified by IPR$^{-1}$, is plotted versus (a) particle velocity and (b) filament length. Black dots: mean values from $200$ configurations, with error bars indicating standard deviation. Solid red and dashed blue curves show analytic approximations for the $W\gg L_f$ (Eq.~\ref{eq:Wlong}) and $W \ll L_f$ (Eq.~\ref{eq:Wshort}), respectively.
    }
    \label{fig:width}
\end{figure}

The width $W$ of the deepest well in the energy landscape can be found by computing the average number of steps away from the global minimum where the landscape first passes a threshold value $C$ above that minimum (Fig.~\ref{fig:range}, inset). In the limit where $L_f \ll W \ll L_\text{dom}$, this width is found through treating the landscape as an entropically biased random walk around the global minimum, yielding the estimate~\cite{supplemental}:

\begin{equation}
    \begin{split}
        W = \frac{1}{3} L_f \left[\left(\frac{C}{s}\right)^2 + 2 \left(\frac{C}{s}\right)\right].
        \label{eq:Wlong}
    \end{split}
\end{equation}
In the limit where the well is confined within a single correlation length ($W \ll L_f$), an alternate derivation~\cite{supplemental} gives:
\begin{equation}
    \begin{split}
        W = 3^{1/3} L_{f} \left(\frac{C}{2n\sigma} \right)^{2/3}.
    \end{split}
    \label{eq:Wshort}
\end{equation}
The estimated well-width provides a good approximation to the $\text{IPR}^{-1}$ metric for localization, with the narrowest distributions occuring at intermediate velocities and filament lengths (Fig.~\ref{fig:width}).
Very short filaments reduce the magnitude of each P\'eclet step, requiring many such steps to climb out of the well; by contrast, very long filaments expand the spatial distance associated with each correlated step, also increasing the well width.

Overall, these results demonstrate that randomly distributed filament tracks result in localization of intermittently transported cargo, over a broad swath of parameter space. Local fluctuations in net filament orientation give rise to a noisy landscape that can be described as a random walk arising from an accumulation of effective local P\'eclet numbers. Localization occurs when there is a deep global minimum in this landscape. Because the range of the landscape scales with the size of the domain, while the well-width surrounding the global minimum does not, particles in a long domain will inevitably become trapped within a highly localized region.

Prior mathematical work highlighted the tendency towards trapping and ultra-slow transport on noisy one-dimensional landscapes~\cite{sinai1983limiting,golosov1984localization}. Here we show how processive motion along scattered parallel filaments defines the landscape features, demonstrating that trapping is determined by a universal compound parameter ($\lambda_\text{geom}/\lambda_\text{kin}$) that incorporates both the filament geometry and the transport kinetics.

Model predictions of cargo localization at intermediate run and filament lengths could be tested by {\em in vitro} experiments using reconstituted microtubule arrays~\cite{bergman2018cargo}.
In the cellular context, these results highlight the importance of cytoskeletal filament arrangements for dictating cargo positioning, even in the absence of large-scale polarity. Cells can ensure cargo delivery to a certain region by restructuring a subset of microtubules so that their polarities converge into a local trap, an observed mechanism that concentrates vesicles at axonal injury sites~\cite{erez2007formation}.

A key implication is that unpolarized random filament arrays are insufficient to ensure broad cargo distribution in proximal dendrites. Localized cargo distributions are expected to arise even without coherent filament organization.
Because the landscape depth scales linearly with transport bias, even cargos that exhibit bidirectional motion (eg: through a combination of kinesin and dynein-driven transport~\cite{hancock2014bidirectional}) will still be subject to trapping if they have a non-zero bias towards either filament end.
However, in live cells a subpopulation of dynamic microtubules may partially smear out the localization effects.
Incorporation of filament dynamics would be a potentially fruitful  direction for further study.

Additional avenues for future work concern the dimensionality of the system. One-dimensional systems are unique in that they inevitably obey detailed balance and can be described via a potential landscape. For parallel filament systems, transverse transport and slow binding / unbinding kinetics would result in splitting into multiple layers of states~\cite{sarpangala2024tunable} with non-trivial memory for transitions between them. Furthermore, other cell types exhibit two- and three-dimensional disordered microtubule networks, where traps have also been  observed in simulations~\cite{ando2015cytoskeletal} and {\em in vitro} systems~\cite{scholz2016cycling}. In such active multidimensional systems, particle localization may involve convergence towards steady-state cycles rather than accumulation in a well.

The random filament system described here serves as an example of biology-inspired physics. Microtubule arrangements observed in mammalian dendrites motivate a quantitative exploration of localization in systems with quenched spatial disorder. The framework presented above illustrates a potentially generalizable approach for linking transport and geometric features to spatial organization in a variety of disordered physical systems.

\section{Acknowledgements}
We are grateful to O. Kogan, G. Huber, E. Tang, and members of the CZI theory group for helpful discussions and insights. Funding was provided by NSF CAREER award \#1848057, NSF grant \#2310229, the UCSD Academic Senate, and the CZI Theory Institute Without Walls.

\section{Data Availability Statement}
Code for simulating particle motion on random filament arrays and for analytic calculation of the steady-state distribution is provided at: \url{https://github.com/lenafabr/transportDisorderedFilamentArray1D}.

\bibliographystyle{unsrt}
\bibliography{collection}

\begin{thebibliography}{10}

\bibitem{hughes1996random}
Barry~D Hughes.
\newblock {\em Random walks and random environments}.
\newblock Oxford University Press, 1996.

\bibitem{debenedetti2001supercooled}
Pablo~G Debenedetti and Frank~H Stillinger.
\newblock Supercooled liquids and the glass transition.
\newblock {\em Nature}, 410(6825):259--267, 2001.

\bibitem{bouchaud1990anomalous}
Jean-Philippe Bouchaud and Antoine Georges.
\newblock Anomalous diffusion in disordered media: statistical mechanisms,
  models and physical applications.
\newblock {\em Phys Rep}, 195(4-5):127--293, 1990.

\bibitem{destrian2026cytoplasmic}
Olivier Destrian, Nicolas Moisan, Ren{\'e}-Marc M{\`e}ge, Benoit Ladoux, Benoit
  Goyeau, and Morgan Chabanon.
\newblock Cytoplasmic crowding acts as a porous medium reducing macromolecule
  diffusion.
\newblock {\em P Natl Acad Sci}, 123(4):e2519599123, 2026.

\bibitem{lu2021search}
Qiao Lu, Deepak Bhat, Darya Stepanenko, and Simone Pigolotti.
\newblock Search and localization dynamics of the crispr-cas9 system.
\newblock {\em Phys Rev Lett}, 127(20):208102, 2021.

\bibitem{chepizhko2013diffusion}
Oleksandr Chepizhko and Fernando Peruani.
\newblock Diffusion, subdiffusion, and trapping of active particles in
  heterogeneous media.
\newblock {\em Phys. Rev. Lett.}, 111:160604, Oct 2013.

\bibitem{sinai1983limiting}
Ya~G Sinai.
\newblock The limiting behavior of a one-dimensional random walk in a random
  medium.
\newblock {\em Theory of Probability \& Its Applications}, 27(2):256--268,
  1983.

\bibitem{goltsev2012localization}
Alexander~V Goltsev, Sergey~N Dorogovtsev, Joao~G Oliveira, and Jose~FF Mendes.
\newblock Localization and spreading of diseases in complex networks.
\newblock {\em Phys Rev Lett}, 109(12):128702, 2012.

\bibitem{chaudhuri2014diversity}
Rishidev Chaudhuri, Alberto Bernacchia, and Xiao-Jing Wang.
\newblock A diversity of localized timescales in network activity.
\newblock {\em {elife}}, 3:e01239, 2014.

\bibitem{zhang2017suppression}
Benjamin~J Zhang, Maysamreza Chamanzar, and Mohammad-Reza Alam.
\newblock Suppression of epileptic seizures via anderson localization.
\newblock {\em J Roy Soc Interface}, 14(127), 2017.

\bibitem{nelson1998non}
David~R Nelson and Nadav~M Shnerb.
\newblock Non-hermitian localization and population biology.
\newblock {\em Phys Rev E}, 58(2):1383, 1998.

\bibitem{wei2020susceptible}
Zong-Wen Wei and Bing-Hong Wang.
\newblock Susceptible-infected-susceptible model on networks with eigenvector
  localization.
\newblock {\em Phys Rev E}, 101(4):042310, 2020.

\bibitem{van2016right}
Petra van Bergeijk, Casper~C Hoogenraad, and Lukas~C Kapitein.
\newblock Right time, right place: probing the functions of organelle
  positioning.
\newblock {\em Trends Cell Biol}, 26(2):121--134, 2016.

\bibitem{misgeld2017mitostasis}
Thomas Misgeld and Thomas~L Schwarz.
\newblock Mitostasis in neurons: maintaining mitochondria in an extended
  cellular architecture.
\newblock {\em Neuron}, 96(3):651--666, 2017.

\bibitem{burute2019cellular}
Mithila Burute and Lukas~C Kapitein.
\newblock Cellular logistics: unraveling the interplay between microtubule
  organization and intracellular transport.
\newblock {\em Annu Rev Cell Dev Bi}, 35:29--54, 2019.

\bibitem{kapitein2011way}
Lukas~C Kapitein and Casper~C Hoogenraad.
\newblock Which way to go? cytoskeletal organization and polarized transport in
  neurons.
\newblock {\em Mol Cell Neurosci}, 46(1):9--20, 2011.

\bibitem{koslover2024searching}
Elena~F Koslover.
\newblock Searching through cellular landscapes.
\newblock In {\em Target Search Problems}, pages 541--577. Springer, 2024.

\bibitem{maza2019intrinsic}
Nycole~A Maza, William~E Schiesser, and Peter~D Calvert.
\newblock An intrinsic compartmentalization code for peripheral membrane
  proteins in photoreceptor neurons.
\newblock {\em J Cell Biol}, 218(11):3753--3772, 2019.

\bibitem{illukkumbura2020patterning}
Rukshala Illukkumbura, Tom Bland, and Nathan~W Goehring.
\newblock Patterning and polarization of cells by intracellular flows.
\newblock {\em Curr Opin Cell Biol}, 62:123--134, 2020.

\bibitem{yap2022spatial}
Chan~Choo Yap and Bettina Winckler.
\newblock Spatial regulation of endosomes in growing dendrites.
\newblock {\em Dev Biol}, 486:5--14, 2022.

\bibitem{chen2015memoryless}
Kejia Chen, Bo~Wang, and Steve Granick.
\newblock Memoryless self-reinforcing directionality in endosomal active
  transport within living cells.
\newblock {\em Nat Mater}, 14(6):589--593, 2015.

\bibitem{goo2017activity}
Marisa~S Goo, Laura Sancho, Natalia Slepak, Daniela Boassa, Thomas~J Deerinck,
  Mark~H Ellisman, Brenda~L Bloodgood, and Gentry~N Patrick.
\newblock Activity-dependent trafficking of lysosomes in dendrites and
  dendritic spines.
\newblock {\em J Cell Biol}, 216(8):2499--2513, 2017.

\bibitem{balint2013correlative}
{\v{S}}tefan B{\'a}lint, Ione Verdeny~Vilanova, {\'A}ngel Sandoval~{\'A}lvarez,
  and Melike Lakadamyali.
\newblock Correlative live-cell and superresolution microscopy reveals cargo
  transport dynamics at microtubule intersections.
\newblock {\em P Natl Acad Sci}, 110(9):3375--3380, 2013.

\bibitem{yogev2016microtubule}
Shaul Yogev, Roshni Cooper, Richard Fetter, Mark Horowitz, and Kang Shen.
\newblock Microtubule organization determines axonal transport dynamics.
\newblock {\em Neuron}, 92(2):449--460, 2016.

\bibitem{richard2019active}
Mathieu Richard, Carles Blanch-Mercader, Hajer Ennomani, Wenxiang Cao,
  Enrique~M De~La~Cruz, Jean-Fran{\c{c}}ois Joanny, Frank J{\"u}licher, Laurent
  Blanchoin, and Pascal Martin.
\newblock Active cargo positioning in antiparallel transport networks.
\newblock {\em P Natl Acad Sci}, 116(30):14835--14842, 2019.

\bibitem{scholz2016cycling}
Monika Scholz, Stanislav Burov, Kimberly~L Weirich, Bj{\"o}rn~J Scholz, SM~Ali
  Tabei, Margaret~L Gardel, and Aaron~R Dinner.
\newblock Cycling state that can lead to glassy dynamics in intracellular
  transport.
\newblock {\em Phys Rev X}, 6(1):011037, 2016.

\bibitem{sarpangala2024tunable}
Niranjan Sarpangala, Brooke Randell, Ajay Gopinathan, and Oleg Kogan.
\newblock Tunable intracellular transport on converging microtubule
  morphologies.
\newblock {\em Biophysical Reports}, 4(3), 2024.

\bibitem{ando2015cytoskeletal}
David Ando, Nickolay Korabel, Kerwyn~Casey Huang, and Ajay Gopinathan.
\newblock Cytoskeletal network morphology regulates intracellular transport
  dynamics.
\newblock {\em Biophys J}, 109(8):1574--1582, 2015.

\bibitem{mlynarczyk2019first}
Paul~J Mlynarczyk and Steven~M Abel.
\newblock First passage of molecular motors on networks of cytoskeletal
  filaments.
\newblock {\em Phys Rev E}, 99(2):022406, 2019.

\bibitem{teryoshin2026encounter}
Lizzy Teryoshin, Mario Hidalgo-Soria, and Elena~F Koslover.
\newblock Encounter times of intermittently running particles.
\newblock {\em Phys Biol}, 2026.

\bibitem{kapitein2010mixed}
Lukas~C Kapitein, Max~A Schlager, Marijn Kuijpers, Phebe~S Wulf, Myrrhe van
  Spronsen, Frederick~C MacKintosh, and Casper~C Hoogenraad.
\newblock Mixed microtubules steer dynein-driven cargo transport into
  dendrites.
\newblock {\em Curr Biol}, 20(4):290--299, 2010.

\bibitem{katrukha2021quantitative}
Eugene~A Katrukha, Daphne Jurriens, Desiree M~Salas Pastene, and Lukas~C
  Kapitein.
\newblock Quantitative mapping of dense microtubule arrays in mammalian
  neurons.
\newblock {\em {elife}}, 10:e67925, 2021.

\bibitem{baas2016stability}
Peter~W Baas, Anand~N Rao, Andrew~J Matamoros, and Lanfranco Leo.
\newblock Stability properties of neuronal microtubules.
\newblock {\em Cytoskeleton}, 73(9):442--460, 2016.

\bibitem{golosov1984localization}
AO~Golosov.
\newblock Localization of random walks in one-dimensional random environments.
\newblock {\em Commun Math Phys}, 92(4):491--506, 1984.

\bibitem{supplemental}
See Supplemental Material for derivations and model details.

\bibitem{nelson2024nonreciprocity}
Aleksandra Nelson and Evelyn Tang.
\newblock Nonreciprocity is necessary for robust dimensional reduction and
  strong responses in stochastic topological systems.
\newblock {\em Phys Rev B}, 110(15):155116, 2024.

\bibitem{zwanzig1988diffusion}
Robert Zwanzig.
\newblock Diffusion in a rough potential.
\newblock {\em P Natl Acad Sci}, 85(7):2029--2030, 1988.

\bibitem{bergman2018cargo}
Jared~P Bergman, Matthew~J Bovyn, Florence~F Doval, Abhimanyu Sharma, Manasa~V
  Gudheti, Steven~P Gross, Jun~F Allard, and Michael~D Vershinin.
\newblock Cargo navigation across 3d microtubule intersections.
\newblock {\em P Natl Acad Sci}, 115(3):537--542, 2018.

\bibitem{erez2007formation}
Hadas Erez, Guy Malkinson, Masha Prager-Khoutorsky, Chris~I De~Zeeuw, Casper~C
  Hoogenraad, and Micha~E Spira.
\newblock Formation of microtubule-based traps controls the sorting and
  concentration of vesicles to restricted sites of regenerating neurons after
  axotomy.
\newblock {\em J Cell Biol}, 176(4):497--507, 2007.

\bibitem{hancock2014bidirectional}
William~O Hancock.
\newblock Bidirectional cargo transport: moving beyond tug of war.
\newblock {\em Nat Rev Mol Cell Biol}, 15(9):615, 2014.

\end{thebibliography}

\end{document}